\documentclass[%
 reprint,
 amsmath,amssymb,
 aps,
]{revtex4-2}

\usepackage[colorlinks=true, allcolors=blue]{hyperref}
\usepackage{graphicx}
\usepackage{dcolumn}
\usepackage{bm}

\newcommand{\rc}[1]{\textcolor{black}{#1}}
\newcommand{\rrc}[1]{\textcolor{black}{#1}}

\newcommand{\gc}[1]{\textcolor{black}{#1}}

\begin{document}

\preprint{APS/123-QED}

\title{\rc{Generation of ultra-relativistic monoenergetic electron bunches via a synergistic interaction of longitudinal electric and magnetic fields of a twisted laser}}

\author{Yin Shi$^{1, 2}$}
\email{shiyin@ustc.edu.cn}
\author{David Blackman$^{1}$}
\author{Dan Stutman$^{3, 4}$}
\author{Alexey Arefiev$^{1}$}%
\email{aarefiev@eng.ucsd.edu}
\affiliation{$^{1}$Department of Mechanical and Aerospace Engineering, University of California at San Diego, La Jolla, CA 92093, USA}%
\affiliation{$^{2}$School of Nuclear Science and Technology, University of Science and Technology of China, Jinzhai Road 96, Hefei 230026, China}%
\affiliation{$^{3}$Extreme Light Infrastructure-Nuclear Physics (ELI-NP)/Horia Hulubei National Institute of Physics and Nuclear Engineering, 077125 Bucharest-Magurele, Romania}
\affiliation{$^{4}$Department of Physics and Astronomy, Johns Hopkins University, Baltimore, Maryland 21218, USA}%

\date{\today}

\begin{abstract}
\rc{We use 3D simulations to demonstrate that high-quality ultra-relativistic electron bunches can be generated upon reflection of a twisted laser beam off a plasma mirror. The unique topology of the beam with a twist index $|l| = 1$ creates an accelerating structure dominated by longitudinal laser electric and magnetic fields in the near-axis region. We show that the magnetic field is essential for creating a train of dense mono-energetic bunches. For a 6.8~PW laser, the energy reaches 1.6~GeV with a spread of 5.5\%. The bunch duration is 320 as, its charge is 60~pC and density is $\sim 10^{27}$~m$^{-3}$. The results are confirmed by an analytical model for the electron energy gain. These results enable development of novel laser-driven accelerators at multi-PW laser facilities.}
\end{abstract}

\maketitle
Multi-PW laser beams have become a reality due to technological developments~\cite{Danson2019,  Lureau2020}, which opens the possibility of achieving laser intensities well in excess of $10^{22}$~W/cm$^2$. There is a push to break the 100~PW limit over the next decade at the Shanghai Superintense-Ultrafast Laser Facility~\cite{Shen2018}. Motivated by these prospects, computational studies have focused on leveraging high-power high-intensity lasers to develop novel particle~(see \cite{Bulanov_POP_2016} and refs. therein) and radiation sources~\cite{Nakamura_PRL_2012, Ridgers_PRL_2012, Ji2014, Stark_PRL_2016, Tao2020, Capdessus2018} for multidisciplinary applications~\cite{Bulanov_2014, Weeks_1997}.

Concurrently with the power increase, new optical techniques are being developed that enable changes to key characteristics of conventional laser beams, thus relaxing what was previously perceived as fundamental constraints. Two striking examples include adjusting the focal-spot velocity~\cite{Froula2018} and twisting the wave-fronts~\cite{Leblanc2017, shi2014}. Simulations show that adjustments to the focal-spot dynamics can drastically improve the performance of laser wakefield accelerators by preventing electrons from outrunning the wake~\cite{Palastro2020}. A wave-front twist that can be achieved via reflection off a fan-like structure for a high-intensity laser pulse~\cite{shi2014} alters the topology of the laser fields. Some of the resulting changes to laser-plasma interactions have been examined in simulations~\cite{vieira2016, Zhang2015, vieira2018, Shi2018,Longman2017,Zhu2019} and experiments~\cite{ Leblanc2017,Denoeud2017}. This work focuses on improvements to electron acceleration. 

Two common approaches to laser-driven electron acceleration are; laser wakefield acceleration~\cite{Esarey2009} that utilizes plasma electric fields, and direct laser acceleration~\cite{gibbon2004short} that relies on the fields of the laser for the energy transfer. The latter can be realized in a plasma (e.g. see \cite{Arefiev_POP_2016}) or in vacuum~\cite{Stupakov_PRL_2001}. \rc{In a conventional laser beam, the energy transfer is dominated by the transverse electric field, $E_{\perp}$. The rate scales as $v_{\perp} E_{\perp} \approx c \sin \theta E_{\perp}$ for ultra-relativistic electrons, making the mechanism less efficient as the electron beam divergence becomes small ($\theta \ll 1$), where $\theta \equiv \arctan(v_{\bot}/v_{\parallel})$ is the angle between the electron velocity $v$ and the laser axis. Another key feature is the transverse electron expulsion caused by electron oscillations in the beam. Inside a plasma, the expulsion can be mitigated by plasma electric and magnetic fields~\cite{Pukhov_1999, Willingale_PRL_2011,gong.pre.2020}. In vacuum, there are no forces to counteract the expulsion, but it can be delayed by injecting electrons with a relativistic longitudinal momentum. Such a configuration can be realized upon reflection of a high-intensity laser pulse off a plasma mirror~\cite{Thevenet2016,Thevenet2016a, Bocoum2016}, as confirmed by experiments with conventional pulses~\cite{Thevenet2016,Bocoum2016}. Electron acceleration in vacuum, also refereed to as Vacuum Laser Acceleration, can further benefit from changes in laser polarization. In a radially polarized beam~\cite{Zaim2017}, $E_\parallel$ dominates near the central axis, so the injected electrons can remain well-collimated while efficiently gaining energy at the rate $\propto v_\parallel E_\parallel \approx c \cos \theta E_{\parallel}$. }

\rc{In this Letter, we show that a high-intensity laser pulse with a properly chosen wave-front twist simultaneously addresses two outstanding problems of direct laser acceleration in vacuum: acceleration efficiency and transverse electron confinement. The twist creates a unique accelerating structure dominated by \emph{longitudinal laser electric and magnetic fields} in the near-axis region. Electron bunches are injected into the laser beam upon its reflection off a plasma mirror. Our three-dimensional particle-in-cell (3D PIC) simulations for a 6.8 PW laser pulse show that the longitudinal laser magnetic field, whose amplitude reaches $\sim 0.2$~MT, is instrumental for generating and sustaining solid electron bunches close to the axis. The confinement enables efficient electron acceleration by $E_\parallel$ that produces 1.6~GeV 60~pC mono-energetic bunches (5.5\% energy spread) with a duration of $\sim 300$~as and a remarkably low divergence of 20~mrad.} These bunches can find applications in research and technology, with one example being free-electron lasers~\cite{Huang2012}.

\begin{figure}
    \centering
    \includegraphics[width=0.99\linewidth]{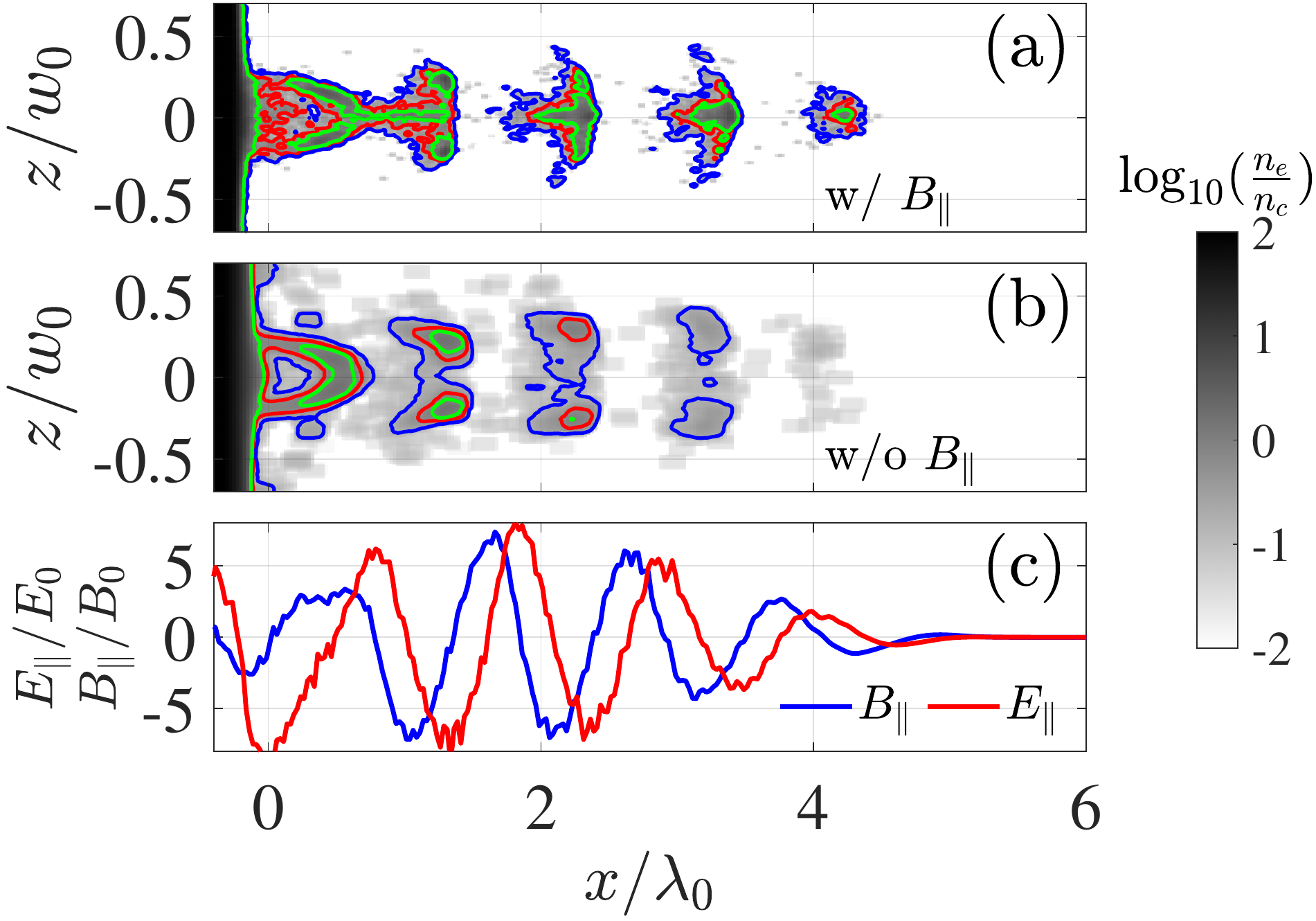}  
\caption{\rc{Electron injection into a reflected laser beam with twisted wave-fronts in 3D PIC simulations with (a) and without (b) $B_x$ in the electron equations of motion.  The density is shown on a log scale at $t = 9$~fs. The blue, red, and green contours denote $n_e = 0.1n_{c}$, 0.5$n_{c}$, and $n_{c}$, where $n_c$ is the critical density. The profiles of longitudinal electric, $E_{\parallel}/E_0$, and magnetic, $B_{\parallel}/B_0$, fields on axis are shown in (c), where $E_0 = B_0 \equiv m_e c \omega/|e|$.}} \label{einject}
\end{figure}

It follows from the paraxial wave equation~\cite{Loudon2003} that the radial field structure of a linearly polarized laser beam with a twist index $l$ propagating along the $x$-axis is given by
\begin{eqnarray}
    && E_{\perp} = B_\perp \propto   \widetilde{r}^{|l|} \exp(-  \widetilde{r}^2 f) \exp(i l \phi) , \\
    && E_{\parallel}^{\pm} = \frac{i \theta_d}{2} \left[ \frac{|l|}{ \widetilde{r}} e^{\mp i \phi} - 2 f  \widetilde{r} \cos \phi \right] E_\perp, \\
    && B_{\parallel}^{\pm} = \frac{\theta_d}{2} \left[ \mp \frac{|l|}{\widetilde{r}} e^{\mp i \phi} - 2i f\widetilde{r} \sin \phi \right] E_\perp ,
\end{eqnarray}
where $ \widetilde{r} = \sqrt{ \widetilde{y}^2 +  \widetilde{z}^2}$, $\phi = \arctan ( \widetilde{z}/ \widetilde{y})$, and $f \equiv (1 - i  \widetilde{x}) / ( 1 +  \widetilde{x}^2)$. The superscripts for $E_\parallel^{\pm}$ and $B_\parallel^{\pm}$ are $\mbox{sgn}(l)$. The transverse coordinates, $\widetilde{y} = y/w_0$ and $\widetilde{z} = z/w_0$, are normalized to the beam waist $w_0$; the longitudinal coordinate, $\widetilde{x} = x/x_R$, is normalized to the Rayleigh range $x_R \equiv \pi w_0^2 / \lambda_0$, where $\lambda_0$ is the laser wavelength. The diffraction angle $\theta_d = w_0/x_R$ is assumed to be small.

\rc{When considering the field structure at $\widetilde{r} \rightarrow 0$ the twist index $l$ is important in determining the field topology. There are three distinct cases: conventional beams where $l = 0$, the first twisted mode with $|l| = 1$, and higher twisted modes where $|l| \geq 2$. At $\widetilde{r} \rightarrow 0$ for $l = 0$, $E_{\parallel}$ and $B_{\parallel}$ vanish while $E_{\perp}$ and $B_{\perp}$ reach their highest amplitude. In the case of $|l| = 1$, we have the opposite: $E_{\parallel}$ and $B_{\parallel}$ reach their highest amplitude at $\widetilde{r} \rightarrow 0$, whereas $E_{\perp}$ and $B_{\perp}$ vanish. In the case of $|l| \geq 2$, all laser fields vanish at $\widetilde{r} \rightarrow 0$. These results are readily generalized for circularly polarized lasers, where $E_z = i \sigma E_y$ and $\sigma = \pm 1$ for the right and left circular polarizations. For $|l| = 1$ and $\sigma = - l$, the longitudinal fields again reach their highest amplitude at $\widetilde{r} \rightarrow 0$, whereas the transverse fields vanish.}

\rc{The near-axis field structure at $|l| = 1$ offers an accelerating configuration where \emph{both} $E_\parallel$ and $B_\parallel$ dominate (contrast this with radially polarized beams that only have $E_\parallel$~\cite{Zaim2017}).} We find that for the circularly polarized beam 
\begin{equation} \label{a and P}
    a_{\parallel} \equiv \frac{|e| E_{\parallel}^{\max}}{m_e c \omega} = \frac{|e| B_{\parallel}^{\max}}{m_e c \omega} \approx 71 \left( \frac{\lambda_0}{w_0} \right)^2 P^{1/2} [\mbox{PW}],
\end{equation}
where $P$ is the period-averaged peak laser power, $e$ and $m_e$ are the electron charge and mass, $c$ is the speed of light, and $\omega$ is the laser frequency. The longitudinal fields are strong even for $\theta_d \ll 1$. We have $E_{\parallel}^{\max} /E_{\perp}^{\max} = B_{\parallel}^{\max} /B_{\perp}^{\max} \approx 0.33 E_{\perp}^{\max}$ for a circularly polarized beam with $w_0 = 3.0~\mu$m and $\lambda_0 = 0.8~\mu$m, which corresponds to $\theta_d \approx 8.5 \times 10^{-2}$. At $P = 6.8$~PW, we have $a_{\parallel} \approx 12.8$, so that $E_{\parallel}^{\max} \approx 5.1 \times 10^{13}$~V/m and $B_{\parallel}^{\max} \approx 170$~kT.

In order to investigate electron acceleration by such a laser beam during reflection off a plasma mirror, we performed 3D PIC simulations using EPOCH~\cite{Arber2015}. \rc{Fig.~\ref{einject}(a) shows the electron density profile, $n_e$, during reflection of a laser beam incident normal to the surface from the right. The reflected beam travels in the positive direction along the $x$-axis.} The target is initialized as a fully ionized hydrocarbon plasma with $n_e = 500 n_c \exp( - 20 (x + 0.3\mu m)/ \lambda_0)$, where $n_c$ = 1.8$\times 10^{27}$~m$^{-3}$ is the critical density for $\lambda_0 = 0.8~\mu$m. The incident pulse has a $\sin^2$ shape with a total duration of 20~fs. Its focal plane is at $x = 0~\mu$m. Simulations with mobile and immobile ions yield identical results, indicating that ion mobility has no impact. See the Supplemental Material for additional details. \rrc{Simulations were independently performed with Smilei~\cite{DEROUILLAT2018351} and EPOCH PIC codes with good agreement. The results were verified with a convergence study performed using Smilei.}

\rc{Fig.~\ref{einject} illustrates the importance of $B_\parallel$ on the formation of solid bunches during the injection process by $E_\parallel$ that takes place during the reflection. The longitudinal field profiles} \rc{along the axis are shown in Fig.~\ref{einject}(c). We performed an additional simulation where we set $B_\parallel = 0$ in the electron equations of motion. \gc{The result from this simulation}, shown in Fig.~\ref{einject}(b), differs from the result of the original simulation, shown in Fig.~\ref{einject}(a). Without the influence from $B_\parallel$, the bunches are hollow, similarly to what is observed for radially polarized beams~\cite{Zaim2017}. They gradually expand, such that $n_e$ on axis becomes very low. Due to the magnetic field, the density in the bunches remains well above $n_c$.} 
 
\gc{The extraction of electrons from the target takes place when the total $E_\parallel$, meaning the sum of the laser field and that from charge separation, is negative. This condition is satisfied for roughly one quarter of each laser period resulting in localized bunches of electrons.}
The upper limit on the density of extracted electrons is estimated as $a_{\parallel} n_c$ and, since $a_{\parallel} \gg 1$, the laser generates dense electron bunches. We find that $n_e > 2n_c$ at the injection stage. The injection process is not sensitive to the temporal laser profile, as confirmed by a simulation with a Gaussian profile. The movie in the Supplemental Material shows the injection process.

\begin{figure}
    \centering
    \includegraphics[width=0.9\linewidth]{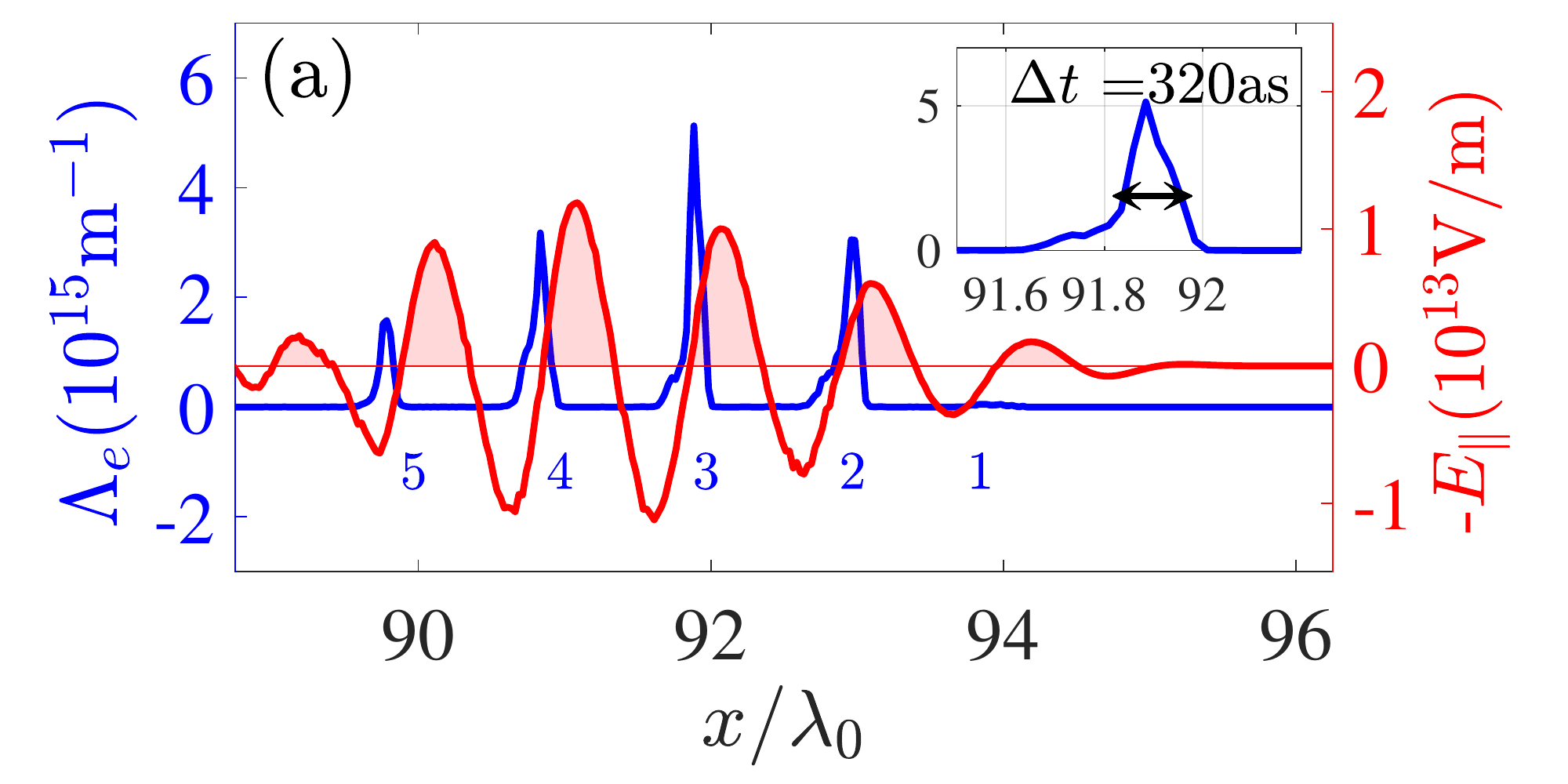} 
    \centering
    \includegraphics[width=0.9\linewidth]{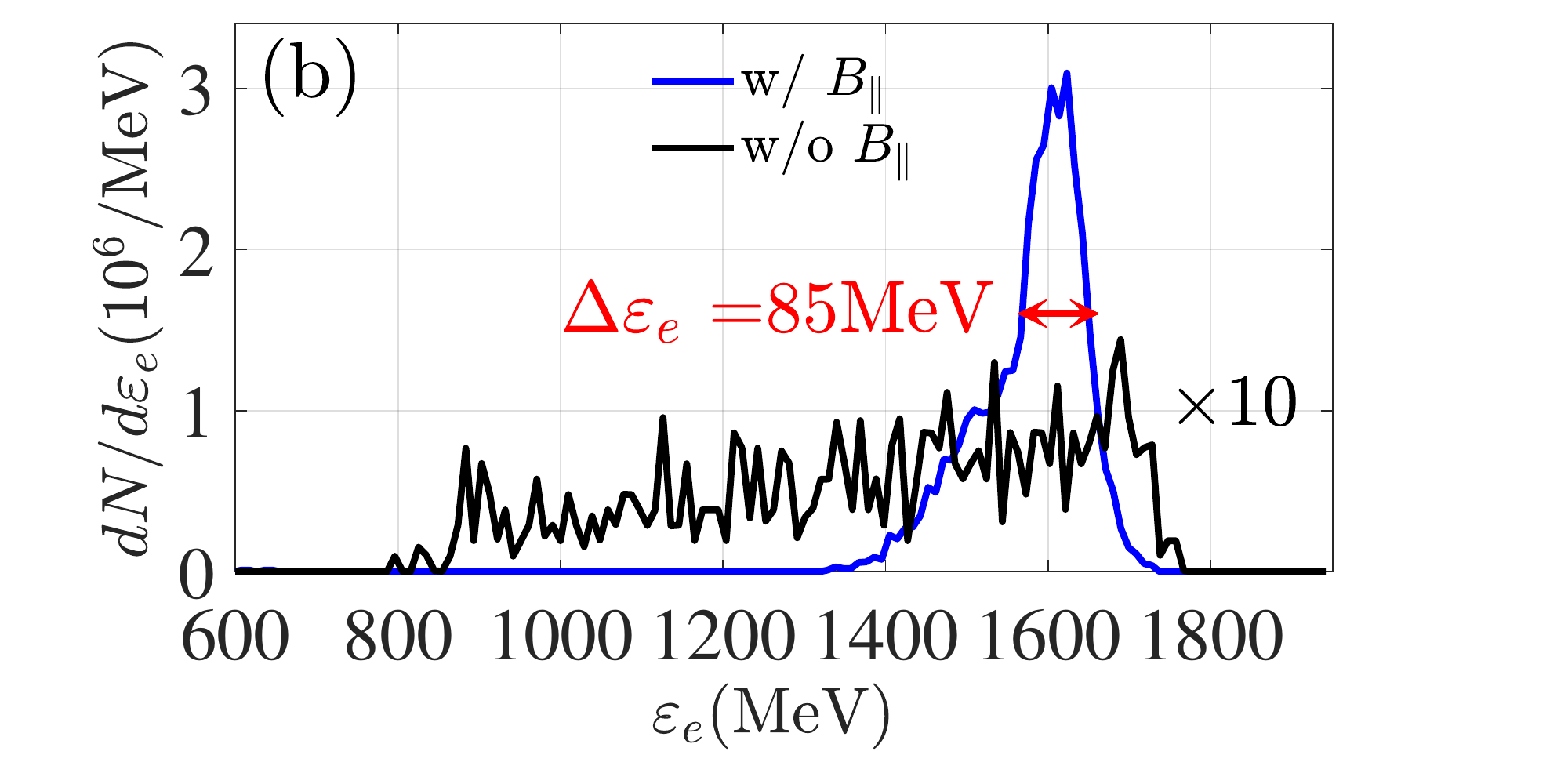} 
\caption{Accelerated electron bunches at $t = 249$~fs. (a)~Linear density $\Lambda_e$, obtained by integrating $n_e$ over $r<1.5$~$\mu$m (blue line, left axis), and the longitudinal electric field $-E_{\parallel}$ (red line, right axis) at $r = 0$. The shading marks the accelerating phase for the electrons. The inset shows the third bunch.  (b)~The energy spectrum $dN/d\varepsilon_{e}$ in the third bunch for electrons with $r<1.5$~$\mu$m. \rc{The black curve is from the simulation with $B_x = 0$ in the electron equations of motion. It was multiplied by 10 to aid the comparison.}} \label{ebunch}
\end{figure}

\begin{figure}
    \centering
    \includegraphics[width=0.49\linewidth]{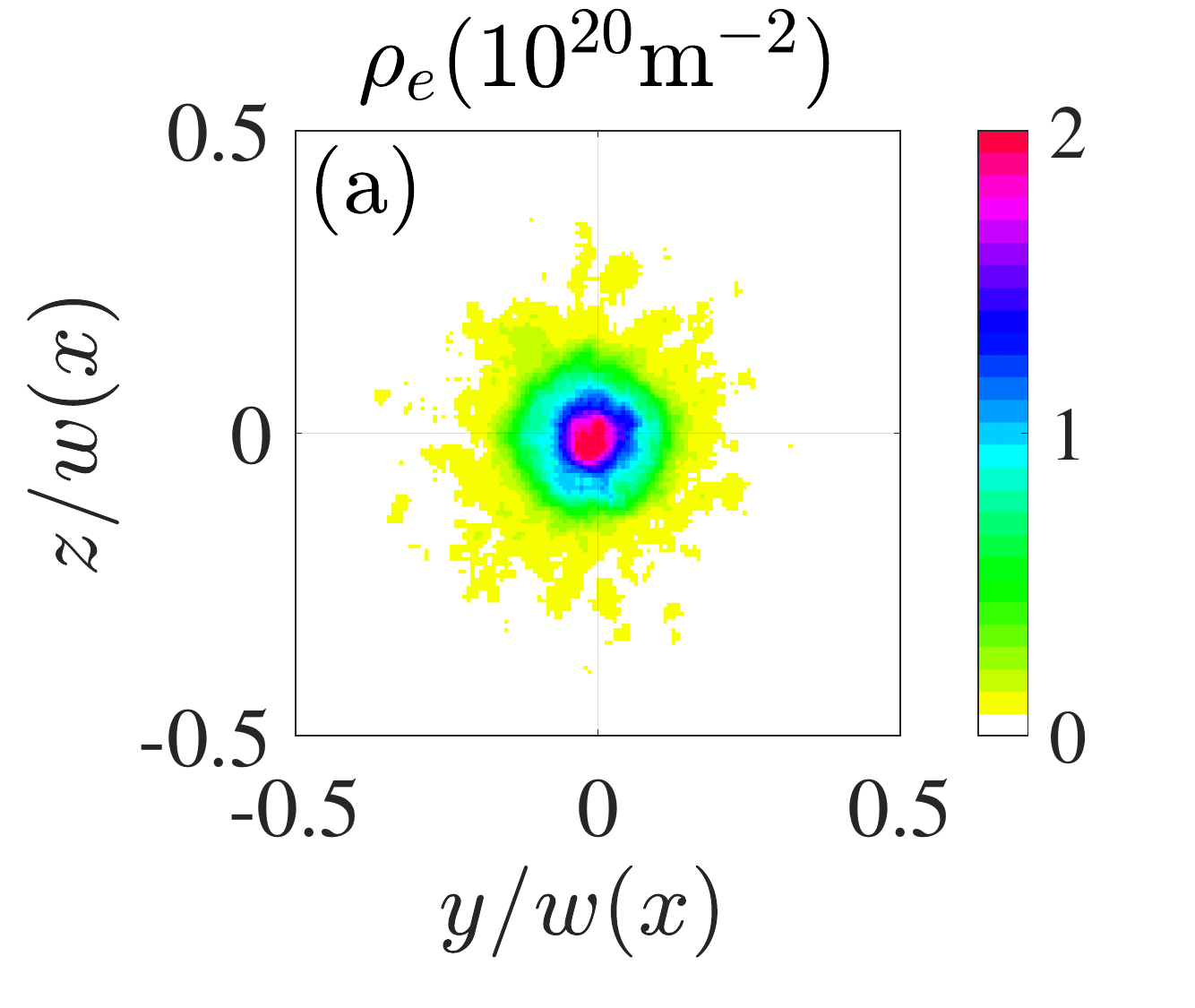} 
    \includegraphics[width=0.49\linewidth]{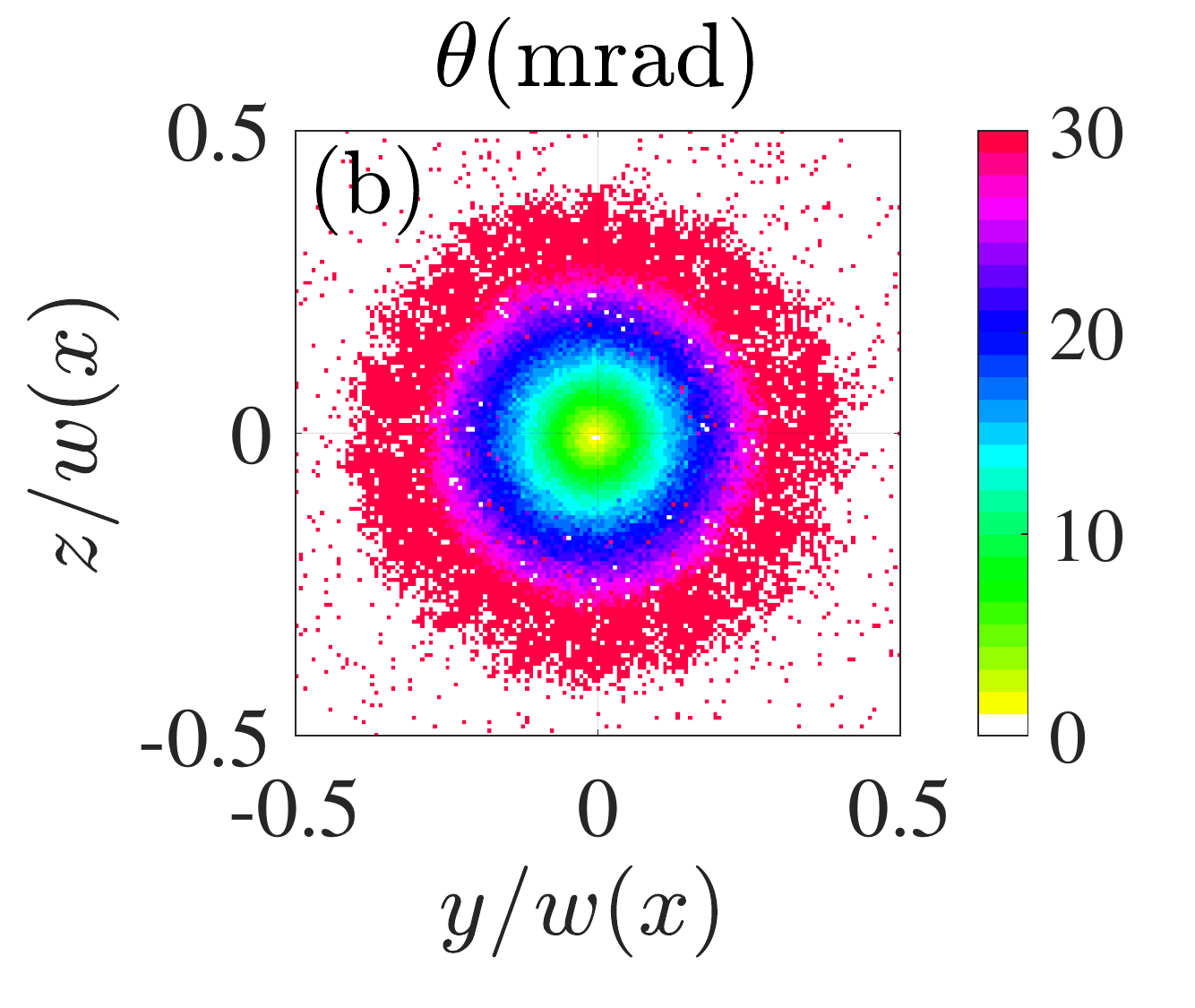} 
\caption{Spatial and angular distribution of the accelerated electrons in the third bunch in Fig.~\ref{ebunch}. (a) The area density, $\rho_e$. (b) The angle of the electron momentum, $\theta \equiv \arctan(p_{\bot}/p_{\parallel})$, with respect to the central axis. The angle is averaged on every mesh cell of the $(y,z)$ plane. \rc{$x$ and $y$ are normalized to the local laser beam width $w(x) = 7.2$ $\mu$m at the bunch location $x = 92 \lambda_0$.}} \label{ebunch_2d}
\end{figure}

\rc{Figs.~\ref{ebunch} and \ref{ebunch_2d} show that the synergistic combination of $E_\parallel$ and $B_\parallel$ produces a train of remarkably dense mono-energetic bunches that are well localized in transverse [Fig.~\ref{ebunch_2d}(a)] and longitudinal [Fig.~\ref{ebunch}(a)] directions. The energy spectrum of the third bunch, shown in Fig.~\ref{ebunch}(b), has a full width at half maximum that is only 5\% of the peak energy that is 1.62~GeV. For the third bunch in the case where $B_\parallel$ is not considered in the equation of motion (black curve) the spectrum is much wider and the number of electrons is ten times lower, emphasizing the importance of $B_\parallel$. The total charge and energy in the third bunch and in all five bunches of the main simulation are 60~pC and 93~mJ and 0.27~nC and 280~mJ.}

\begin{figure} [tbp!]
    \centering
    \includegraphics[width=0.9\linewidth]{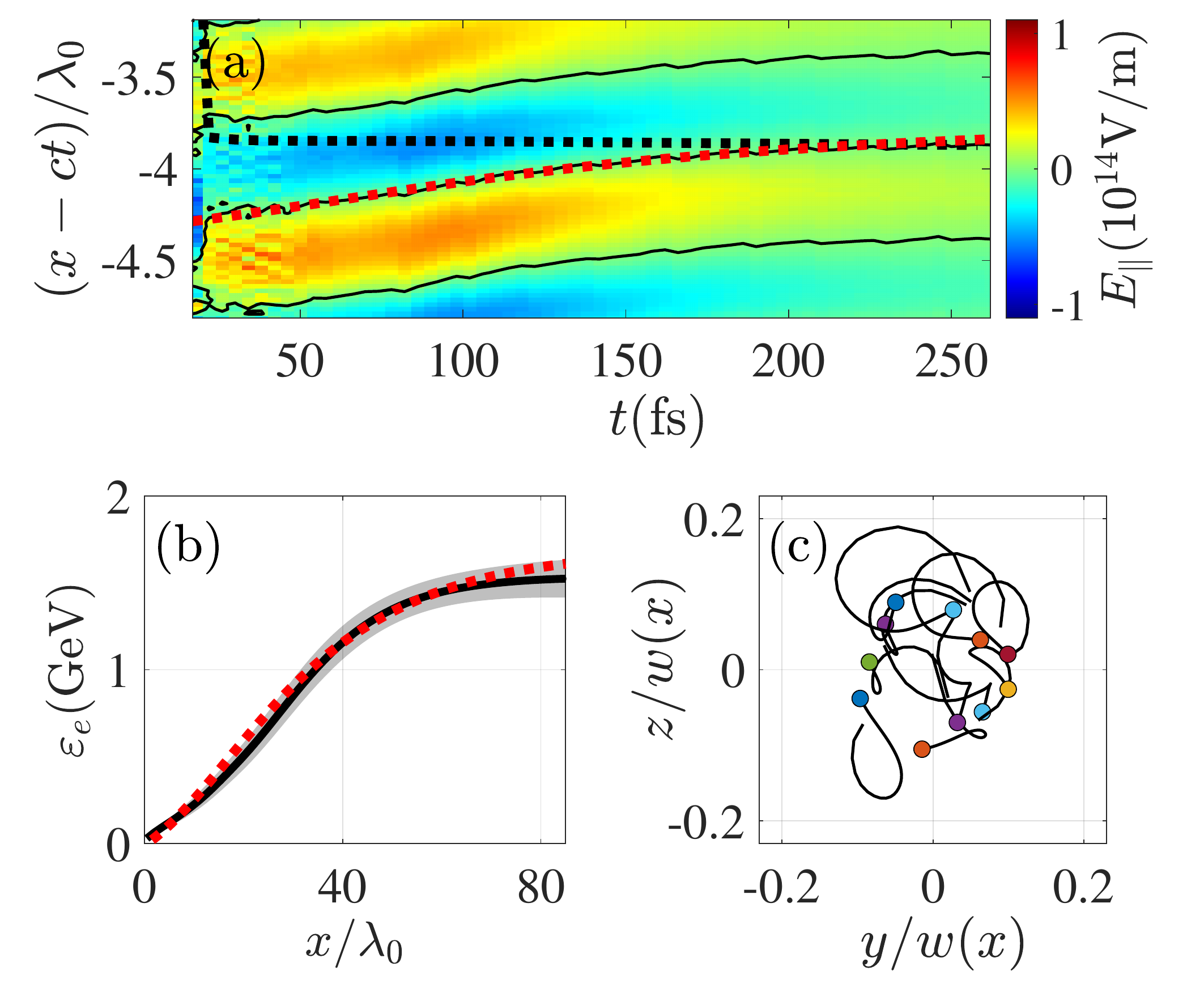} 
\caption{Evolution of the third electron bunch. (a) Longitudinal position of a representative electron (black dotted curve) with respect to the longitudinal electric field $E_{\parallel} (r=0)$. The solid contours denote $E_{\parallel} = 0$. The red dotted curve is the analytical prediction for $v_{ph}$. \rc{(b) Electron energy in the mono-energetic third bunch, with the shaded region showing the standard deviation, vs the position of the bunch, $x$. The red dotted curve is the prediction of our model given by Eq.~(\ref{main_result}) for $\Phi_0 = - 0.18\pi$. (c) Projection of randomly picked electron trajectories from the third bunch onto the beam cross-section. $y$ and $z$ are normalized to the spot size $w(x)$. The markers show electron positions at $t = 249$~fs.} } \label{etrajEx}
\end{figure}

\rc{The laser magnetic field $B_{\parallel}$ provides transverse electron confinement in the region with strong $E_\parallel$, as shown in Fig.~\ref{etrajEx}(c). By preventing electrons from reaching the region with $r \sim w(x)$, where $w(x) = w_0 \sqrt{1 + x^2/x_R^2}$ is the transverse size of the beam, $B_\parallel$ ensures that the electrons are unable to sample strong $E_\perp$. As a result, the divergence angle remains remarkably low.} \rc{In the third bunch, Fig.~\ref{ebunch_2d}(b), it is below 20 mrad for the majority of the electrons.} \rc{The confinement may be particularly helpful in setups with oblique laser incidence where the electrons are injected with increased transverse momentum that would lead to their loss in the absence of $B_\parallel$.} \rc{The importance of $B_\parallel$ can be assessed by estimating the Larmor radius, $r_L$, for an electron injected with transverse relativistic momentum $p_\perp$. It follows from Eq.~(\ref{a and P}) that $r_L/w_0 \approx 2.2 \times 10^{-3} (p_\perp / m_e c) (w_0 / \lambda_0 ) P^{-1/2} [\mbox{PW}]$. For our laser parameters, $r_L < 0.5 w_0$ for $p_\perp < 158 m_e c$. In the presented simulation with normal incidence, $p_{\bot} \approx 20 m_e c$, so $B_\parallel$ provides strong confinement.}

The maximum value of the area density in a bunch, $\rho_e$, (integral of $n_e$ along the bunch) is obtained by taking into account that the injection during one laser period stops once the space-charge of the extracted electrons shields $E_\parallel$ of the laser. This yields $\rho_e \approx a_{\parallel} n_c c / \omega$ or $\rho_e [\mbox{m}^{-2}] \approx 1.3 \times 10^{22} P^{1/2} [\mbox{PW}] \lambda_0 [\mu\mbox{m}] w_0^{-2} [\mu\mbox{m}]$. We find that the normalized emittance in $y$ and $z$ is $\widetilde{\epsilon}_{rms, y} \approx \widetilde{\epsilon}_{rms, z} \approx 1.3 \times 10^{-5}$. The parameters for all five bunches are summarized in the Supplemental Material.

In the considered regime, the accelerated electrons retain most of the energy they gain from the laser field due to the diffraction of the laser beam. Fig.~\ref{etrajEx}(a) shows how a single electron from the third bunch moves with respect to the wave-fronts of $E_{\parallel}$. The longitudinal electron location is shown in a window moving forward with the speed of light. The amplitude of $E_\parallel$ drops appreciably by the time the electron slips into the decelerating phase (shown in yellow and red), so the laser is unable to reduce the electron energy. This observation is confirmed by Fig.~\ref{etrajEx}(b) that shows saturation of the average energy of the electrons in the third bunch. The shaded area shows the standard deviation for the energy in the bunch. A simulation tracking the acceleration process over a longer time (see the Supplemental Material) \rc{definitively confirms the energy retention by the electron bunch}. An additional simulation for a twisted linearly-polarized beam with ($l = -1, \sigma = 0$) shows the robustness of the observed injection and acceleration processes.

\rc{The energy gain can be predicted by accounting for the electron dephasing from $E_\parallel$. In our case, $c - v_x \ll v_{ph} - c$ and the dephasing is primarily dictated by the degree of superluminosity, i.e. by $v_{ph} - c$.} Near the axis we have
\begin{eqnarray}
   && E_{\parallel} =   E_* \sin \left( \Phi + \Phi_0 \right) / (1 + x^2/x_R^2), \label{E_||}\\ 
   && \Phi = 2 \left[ \theta_d^{-2} (x/x_R) - \tan^{-1}  (x/x_R) \right] - \omega t  , \label{phase}
\end{eqnarray}
where $E_* > 0$ is the amplitude of a given wave-front. $\Phi_0$ can be interpreted as the injection phase for the considered bunch that starts its acceleration at $x \approx 0$ at $t \approx 0$. It follows from Eq.~(\ref{phase}) for $\theta_d \ll 1$ that $(v_{ph} - c)/c \approx \theta_d^2 / \left( 1 + x^2 / x_R^2 \right)$. This result, shown as a red dotted curve in Fig.~\ref{etrajEx}(a), describes well the motion of wave-fronts in our 3D PIC simulation. To find the phase slip, we set $x \approx ct$ in Eq.~(\ref{phase}), which yields $\Delta \Phi \approx - 2 \tan^{-1} (x/x_R)$. The phase velocity is superluminal at $x = 0$, but it decreases to $c$ at $x \gg x_R$. As a result, the phase slip is just $\Delta \Phi \approx - \pi$ at $x \gg x_R$.

Taking into account that the phase slip is less than one period at $x = x_R$, we estimate the momentum gain by the electrons as $\Delta p_{\parallel} \approx |e| E_* \Delta t \approx  |e| E_* x_R / c$. To refine this result, we integrate $dp_\parallel / dt = -|e| E_\parallel$ over $t$ with the assumption that $x \approx ct$, which, after change of variables, yields: 
\begin{eqnarray}
    &&\Delta p_{\parallel} \approx -|e| \int_{0}^{x} E_* \frac{ x_R^2}{x_R^2+x^2} \sin (\Delta \Phi + \Phi_0) \frac{dx}{c} \nonumber \\
    && = \frac{|e| E_* x_R}{2c} \left[ \cos \Phi_0 - \cos \left(\Phi_0 - 2 \tan^{-1} \frac{x}{x_R} \right) \right], \label{main_result}
\end{eqnarray} 
where we used Eq.~(\ref{E_||}) and the expression for $\Delta \Phi$. At $x \gg x_R$, we have $\Delta p_{\parallel} \approx |e| E_* \cos (\Phi_0) x_R / c$. 
Fig.~\ref{etrajEx}(b) shows that this model reproduces the energy gain of the third bunch relatively well for $\Phi_0 \approx -0.18 \pi$. 

Our model predicts that the maximum energy gain by an electron bunch scales as
\begin{equation} \label{max energy}
    \varepsilon_{\max} [\mbox{GeV}] \approx 0.72 \cos (\Phi_0) P^{1/2} [\mbox{PW}], 
\end{equation}
where we took into account that for the ultra-relativistic electrons $\varepsilon_{\max} / m_e c^2 \approx \Delta p_{\parallel}/m_e c \approx a_{\parallel} \cos (\Phi_0) \omega x_R / c$. We expressed $a_{\parallel}$ in terms of $P$ using Eq.~(\ref{a and P}). The result is independent of the spot size $w_0$ and wavelength $\lambda_0$. Eq.~(\ref{max energy}) agrees well with the presented 3D PIC simulation results. For 6.8~PW and $\Phi_0 = -0.18 \pi$, we have $\varepsilon_{\max} \approx 1.6$~GeV, which matches the energies in Fig.~\ref{ebunch}(b) and \ref{etrajEx}(b). One may want to use an aperture to extract the electron beam from the laser.\rc{It is important to emphasize that the simplicity of the model owes itself to the presence of $B_\parallel$ that keeps the electrons at $r \ll w(x)$.}

In order to verify the robustness of the described mechanism for generating dense bunches of energetic electrons, we have performed additional 3D PIC simulations. Our observations also hold at lower power of $P = 425$~TW (1/16 of the original power). A single bunch has a flat energy distribution from 50 to 400~MeV, a charge of 14.6~pC, a duration of 400~as, and a normalized transverse emittance of 3.2~$\mu$m. Our mechanism strongly relies on $a_{\parallel} \gg 1$, so it is not surprising that the performance of the mechanism deteriorates with further power decrease.

In summary, we have examined using 3D kinetic simulations an interaction of a twisted laser beam  with a plasma mirror and found that the unique topology of a laser beam with a twist index $|l| = 1$ generates high quality dense mono-energetic electron bunches. \rc{The key feature is the synergistic interplay of longitudinal electric and magnetic fields that dominate the field structure near the axis of the beam.} For a laser with 6.8~PW of incident power, one bunch has an energy of 1.62~GeV, an energy spread of 5\%, a charge of 59.6~pC, a duration of 320~as, and a normalized transverse emittance of 12.9~$\mu$m. Using our scheme, high-power high-intensity state-of-art laser systems and novel optical techniques for creating twisted wave-fronts can be combined to design a source of high-quality dense attosecond bunches of ultra-relativistic electrons for potential applications in research and technology~\cite{Norbert2019, Black2019}. 

This  work  was  supported  by  the NSF  (Grant  No. 1903098). Y.S. acknowledges the support of Newton International Fellows Alumni follow-on funding. D.S. acknowledges support by grant PN-III-P4-ID-PCCF-2016-0164  of the Romanian National Authority for Scientific Research and Innovation.
Simulations were performed with EPOCH (developed under UK EPSRC Grants EP/G054950/1, EP/G056803/1, EP/G055165/1 and EP/ M022463/1) using HPC resources provided by TACC. \rc{We thank J.-L. Vay from LBL for insightful discussions on numerical dispersion.}

\end{document}